\documentclass[12pt]{article}

\usepackage[latin1]{inputenc}
\usepackage{amsmath,cite}
\usepackage{amsfonts}
\usepackage{amssymb,MnSymbol,slashed}
\usepackage{amsxtra}
\usepackage[margin=3cm]{geometry}
\usepackage{color}
\usepackage{hyperref}
\hypersetup{linktocpage=true}
\usepackage{graphicx}
\usepackage{placeins}
\usepackage{upgreek} 
\usepackage{mathrsfs}
\usepackage{accents}
\usepackage{multirow} 
\usepackage{tikz}

\numberwithin{equation}{section}
\newcommand{\beq}{\begin{equation}}
\newcommand{\eeq}{\end{equation}}
\def\be {\begin{equation}}
\def\ee {\end{equation}}
\def\bs#1\es{\begin{split}#1\end{split}}
\def\ba#1\ea{\begin{align}#1\end{align}}
\def\baed#1\eaed{\begin{aligned}#1\end{aligned}}
\def\bged#1\eged{\begin{gathered}#1\end{gathered}}
\def\bea{\begin{eqnarray}}
\def\eea{\end{eqnarray}}

\def\nn{\nonumber}

\def\d{\delta}

\def\e{\epsilon}

\def\F{\Phi}

\def\G{\Gamma}
\def\h{\eta}
\def\k{\kappa}
\def\l{\lambda}

\def\m{\mu}
\def\n{\nu}
\def\o{\omega}
\def\O{\Omega}

\def\r{\rho}

\def\bls{\bigg [}
\def\brs{\bigg ]}

\newcommand{\tb}[1]{ \textrm{\tiny{{({#1})}}} {} }

\def\R{\text{Re}}

\newcommand{\cT}{\mathcal{T}}

\newcommand{\cW}{\mathcal{W}}

\newcommand{\cR}{\mathcal{R}}

\def\cO{{{\mathcal O}}}
\def\cM{\mathcal{M}} 
\def\cN{\mathcal{N}}
\def\cV{\mathcal{V}}
\def\cX{\mathcal{X}}
\def\cZ{\mathcal{Z}}

\def\Re{\text{Re}}

\def\pa{\partial}
\def\na{\nabla}
\def\fr{\frac}
\def\tfr{\tfrac}

\def\we{\wedge}
\def\ra{\rightarrow}

\def\tbzero{{\text{\tiny{(0)}}}}

\def\tbtwo{{\text{\tiny{(2)}}}}
\def\eppr{\alpha}

\newcommand{\wh}[1]{ {\hat{#1}}{} }
\newcommand{\til}[1]{ {\tilde{#1}} }

\let\foo\bar 
\renewcommand{\bar}[1]{ {\foo{  #1} }{} }

\newlength{\dhatheight}

\begin{document}

\baselineskip=16pt
\setlength{\parskip}{6pt}

\begin{titlepage}
\begin{flushright}
\parbox[t]{1.4in}{
\flushright MPP-2015-149}
\end{flushright}

\begin{center}

\vspace*{1.5cm}

{\Large \bf The effective action of warped M-theory reductions }\\[.1cm]
{\Large\bf with higher-derivative terms}\\[.1cm]
{\large \bf -- Part II --}
\end{center}

\vskip 1.5cm
\renewcommand{\thefootnote}{}

\begin{center}
 \normalsize 
 Thomas W. Grimm, Tom G. Pugh, Matthias Weissenbacher \footnote{{ grimm, mweisse, pught @ mpp.mpg.de}}

\vskip 0.5cm
Max Planck Institute for Physics,\\
F\"ohringer Ring 6, 80805 Munich, Germany
\end{center}

\vskip 1.5cm
\renewcommand{\thefootnote}{\arabic{footnote}}

\begin{center} {\bf ABSTRACT } \end{center}

We study the three-dimensional effective action 
obtained by reducing eleven-dimensional 
supergravity with higher-derivative terms 
on a background solution including a warp-factor, an eight-dimensional
compact manifold, and fluxes. 
The dynamical fields are K\"ahler deformations and vectors from 
the M-theory three-form. We show that the 
potential is only induced by fluxes and the naive contributions 
obtained from higher-curvature terms 
on a Calabi-Yau background vanish once the 
back-reaction to the full solution is taken into account. 
For the resulting three-dimensional action we analyse the 
K\"ahler potential and complex coordinates and show 
compatibility with $\cN=2$ supersymmetry.
We argue that the higher-order result is also compatible with 
a no-scale condition. We find that 
the complex coordinates should be formulated as divisor 
integrals for which a non-trivial interplay between the 
warp-factor terms and the higher-curvature terms allow 
a derivation of the moduli space metric. 
This leads us to discuss  
higher-derivative corrections to the M5-brane action. 

\end{titlepage}

\newpage
\noindent\rule{\textwidth}{.1pt}		
\tableofcontents
\vspace{20pt}
\noindent\rule{\textwidth}{.1pt}

\setcounter{page}{1}
\setlength{\parskip}{9pt} 

\newpage

\section{Introduction}

Three-dimensional effective theories arising from M-theory on eight-dimensional 
compact manifolds are of both conceptual 
as well as phenomenological interest. For example M-theory backgrounds that admit a torus fibration allow a lift of the three-dimensional theories to four dimensions \cite{Vafa:1996xn} that can be phenomenologically compelling. 
The four-dimensional theories are minimally supersymmetric admitting four supercharges
if the three-dimensional effective theories are $\cN=2$ supersymmetric. With this 
motivation in mind, we will study in this work a general class of M-theory reductions and argue 
that they are compatible with $\cN=2$ supersymmetry. More precisely, 
we analyse the perturbations of warped solutions with an eight-dimensional 
compact internal manifold and background fluxes. 

The background solutions of interest have first been considered in \cite{Becker:1996gj}.
At leading order the background is 
simply a direct product of three-dimensional Minkowski space and a Calabi-Yau fourfold without background fluxes. 
When including background fluxes and all relevant 
higher-derivative terms it was shown that the internal background is conformally K\"ahler 
with vanishing first Chern class, but that the metric is non-Ricci-flat even when allowing for a conformal
rescaling including the warp factor \cite{Becker:2001pm,Grimm:2014xva}. 
A complete check of supersymmetry at this order of derivatives is still missing.
However, using a proposed correction the eleven-dimensional gravitino variations based on \cite{Lu:2003ze,Lu:2004ng} 
it was shown in \cite{Grimm:2014xva} that supersymmetry can be preserved by this background. 
The fluctuations of this solution were then studied in \cite{Grimm:2014efa} and their three-dimensional 
effective action was derived by dimensional reduction. More precisely, 
a finite number of K\"ahler deformations of the metric and 
vector deformations of the M-theory three-form were included when 
reducing the leading eleven-dimensional supergravity action of 
\cite{Cremmer:1978km} corrected by the terms fourth order in the Riemann
curvature \cite{Duff:1995wd,Green:1997di,Green:1997as,Kiritsis:1997em,Russo:1997mk,Antoniadis:1997eg,Tseytlin:2000sf}, 
and the higher-derivative terms quadratic in the M-theory three-form \cite{Liu:2013dna}. 
The resulting effective action was given to quadratic order in the scale parameter $\alpha \propto \ell_{M}^{3}$, where $\ell_M$ is the 
eleven-dimensional Planck length. The present work is a continuation of \cite{Grimm:2014efa}, which 
discusses in more detail the scalar potential and the supersymmetry properties of the three-dimensional 
effective theory. 

In reference \cite{Grimm:2014efa} it was shown that the number of K\"ahler deformations and vector zero modes are 
still given by the dimension of the second cohomology of the compact manifold. While the Kaluza-Klein reduction is originally no 
longer performed by an expansion into harmonic forms of the underlying Calabi-Yau geometry, it was shown 
that all corrections, associated with these alternative fluctuations, drop from the effective action. The kinetic terms 
for the deformations and vectors in the three-dimensional effective theory 
were written in terms of a single higher-curvature building 
block $Z_{m\bar m n \bar n} = \frac{1}{4!} (\epsilon_{8} \epsilon_{8} R{}^3)_{m\bar m n \bar n}$, where 
$ R$ is the internal Riemann tensor in the underlying Calabi-Yau metric.\footnote{The equivalent 
quantity on a Calabi-Yau threefold was found to be important in \cite{Katmadas:2013mma}.}
Furthermore, in \cite{Grimm:2014efa} the warp-factor was fully included in the reduction. It was shown 
that the effective action contains integrals depending on the bare warp-factor and its first derivatives
with respect to the K\"ahler structure deformations. Remarkably, these derivative couplings only appear through covariant
derivatives under a moduli dependent scaling symmetry under which the K\"ahler structure deformations and the 
warp-factor transform. The classical leading order three-dimensional $\cN=2$ theory
obtained from M-theory on a Calabi-Yau fourfold with background 
fluxes was first found in \cite{Haack:1999zv,Haack:2001jz}, 
while recent derivations of $\cN=1$
effective theories arising from M-theory flux compactifications can be found in \cite{Bonetti:2013fma,Prins:2013wza,Prins:2015nda}.
Let us note that previous works on warped compactifications of 
M-theory and Type IIB  include \cite{Dasgupta:1999ss,Giddings:2001yu,Giddings:2005ff,Burgess:2006mn,Shiu:2008ry,Douglas:2008jx,Martucci:2009sf,Underwood:2010pm,Grimm:2012rg,Frey:2013bha,Martucci:2014ska}. 
Our analysis of the $\cN=2$ supergravity data also extends the works \cite{Grimm:2013gma,Grimm:2013bha,Junghans:2014zla},
 which presented a partial reduction 
 from eleven to three dimensions including some of the relevant higher-derivative terms. Recent interesting 
 results on dimensional reductions with higher-derivative terms can also be found in \cite{Ciupke:2015msa,Minasian:2015bxa}.

In a first step, and as completion of the results of \cite{Grimm:2014efa}, we derive the scalar potential 
for the K\"ahler structure deformations by dimensional reduction. Interestingly, reducing 
the higher-curvature terms on the leading order Calabi-Yau background 
it appears that they become massive with a coupling purely depending on the geometry. 
However, we will show that these mass terms are precisely cancelled by the higher-order
corrections in the solution arising as a back-reaction effect. The remaining scalar potential 
is only induced by background fluxes as in \cite{Haack:2001jz}. This gives a further test that 
the included fluctuations are indeed the relevant light degrees of freedom and highlights 
the interplay from back-reaction effects in the solution and the corrections to the 
effective theory. 

In order to reveal the supersymmetry properties of the three-dimensional effective action 
we discuss its promotion into the standard $\cN=2$ form. In 
three space-time dimensions massless vectors are dual to scalars and the dynamics 
of the light modes therefore should be describable by a K\"ahler potential and a set of complex coordinates. 
We study the order by order expansion of the K\"ahler potential and complex coordinates in 
the K\"ahler structure fluctuations. The coefficients are deduced by comparison with the 
dimensionally reduced action. We infer compatibility with $\cN=2$ supersymmetry and
argue that a no-scale condition can be implemented. Since the dimensional reduction 
only includes the leading-order terms in the fluctuations we are not able to completely 
fix all coefficients by the comparison alone.
The fundamental `all-order' expression, as it is known for the classical reduction without higher-curvature 
terms \cite{Haack:1999zv,Haack:2001jz}, turns out to be even more difficult to find. We argue that this 
problem lies in fixing the complex coordinates and should be approached by 
introducing divisor integrals. These integrals should be matched with  the actions
of M5-branes wrapped on divisors. We make steps towards finding an all order expression 
for the complex coordinates and K\"ahler potential. An intriguing interplay 
between variations of warped divisor integrals and higher-curvature terms 
via the warp-factor equation allows the compatibility with the dimensional 
reduction to be shown. As a byproduct this suggests that the M5-brane action 
should receive higher-curvature corrections that parametrise the non-harmonicity 
of the fourth Chern-form of the background geometry.

The paper is organised as follows. In section \ref{dim_rep} we recall the
background solutions, introduce an appropriate set of fluctuations, and review
the dimensionally reduced effective action following \cite{Grimm:2014efa}. In addition, we 
analyse the scalar potential and comment on a scaling symmetry of the effective action. 
The $\cN=2$ supersymmetric structure and the no-scale condition are discussed in 
section \ref{susy_structure}. We derive the K\"ahler potential and complex coordinates
as an expansion in the fluctuations and later propose a definition using divisor integrals.

\section{Dimensional reduction of the M-theory action} \label{dim_rep}

In this section we first review the background solution of 
eleven-dimensional supergravity including higher-derivative terms following  
\cite{Becker:2001pm,Grimm:2014xva}.
We then introduce the variations of the solutions considered in \cite{Grimm:2014efa}
and show that at order $\alpha^2$ they only admit a scalar potential 
due to background fluxes. We recall the complete three-dimensional 
effective action including all order $\alpha^2$-terms 
following \cite{Grimm:2014efa} and discuss its various building 
blocks and symmetries.

\subsection{Higher-order background solution} \label{background_solution}

To begin with we first review the warped solutions following \cite{Becker:2001pm,Grimm:2014xva}.
These backgrounds satisfy the eleven-dimensional field equations to order $\eppr^2 = \fr{( 4 \pi \k_{11}^2)^\fr23}{(2 \pi)^4 3^2 2^{13}}$. 
The eleven-dimensional metric in this background takes the form
\ba \label{Metric11d}
 d \wh s^2 &= e^{\eppr^2  \F } ( e^{-2 \eppr^2 W}   \h_{\m\n} dx^\m dx^\n  +  2 e^{  \eppr^2 W}  \check g_{m \bar n} dy^m dy^\bar n )\ + \cO(\eppr^3) ,
\ea
where $\h_{\m\n} $ is the three-dimensional Minkowski metric and  
\ba 
\check g_{m \bar n} = g_{m \bar n} + \eppr^2 g^\tbtwo_{m \bar n} + \cO(\eppr^3)\, .
\label{gexp}
\ea
The internal compact manifold will be denoted by $Y_4$ 
and $\F $ and $W$ are scalar functions on this space. $W$ is the warp-factor and is constrained by a differential equation 
\eqref{warpfactoreq}. For 
simplicity, and in contrast to \cite{Grimm:2014xva,Grimm:2014efa}, we will not always indicate the $\eppr$-order at the symbol, i.e.~we
write $\F \equiv \F^\tbtwo $ and $W \equiv W^\tbtwo$. Furthermore, we will use $g_{m \bar n} \equiv g^{\tbzero}_{m \bar n}$
to denote the zeroth-order metric. All quantities, such as higher-curvature terms on $Y_4$, are 
always evaluated in this zeroth-order metric $g_{m \bar n}$ unless indicated explicitly. This will simplify the 
notation compared to \cite{Grimm:2014xva,Grimm:2014efa}. For example, $\F$ represents 
an eleven-dimensional Weyl rescaling and is given in terms of the lowest order metric $g_{m \bar n}$ as 
\ba \label{def-Z}
\F&= - \tfr{512}{3} Z\, , & 
Z &= * ( J \we c_3 )\ ,
\ea
where $*$ is the Hodge-star on $Y_4$ in the metric $g_{m \bar n}$, $J$ is the K\"ahler form 
built from $g_{m\bar n}$ and  
$c_3$ is the third Chern form built from $g_{m \bar n}$ as 
\ba
c_3 = -\fr{i}{3} \cR_{m}{}^n \we \cR_{n}{}^{r} \we \cR_{r}{}^s\ .
\ea
In this expression we have used the definition of the two-form $\cR_{m}{}^n$ that is built from the Riemann tensor as $\cR_r{}^s  = R_{m \bar n r}{}^{s} dy^m dy^{\bar n}$.

In order to give the expressions \eqref{gexp} and \eqref{def-Z} we note that 
at zeroth order in $\alpha$ the background is a direct product and 
$g_{m \bar n}$ is a Ricci flat metric on a Calabi-Yau fourfold. 
We therefore can introduce complex indices, which here and in the following 
always refer to the zeroth order complex structure on the internal manifold. 
On a Calabi-Yau fourfold there exists a nowhere vanishing covariantly constant 
K\"ahler form $J$ and holomorphic $(4,0)$-form $\O$ satisfying
\ba
d J = d \O = 0 \, . 
\ea 
In what follows we will work in conventions in which the internal space indices are 
raised and lowered with the lowest order internal space metric $g_{m \bar n} $.
At second order in $\eppr$ the metric is corrected by $g_{m \bar n}^{\tbtwo}$
in \eqref{gexp}. This is constrained by the higher-derivative Einstein equations that are solved by 
\ba
\label{g2Expression}
g_{m \bar n}^{\tbtwo} &= 768\pa_m \bar \pa_{\bar n} \til F\ , & 
 \til F &= * ( J \we J \we F_4) \ .
\ea
Here $F_4$ is a four-form parameterising the non-harmonic part of the third Chern-from.
Since $c_3$ is closed on a K\"ahler manifold we may write 
\ba \label{def-F4}
c_3 &= H c_3 + i \pa \bar \pa F_4 \, , 
\ea
where $H$ indicates the projection to the harmonic part associated with the metric $g_{m \bar n }$. 
The expression \eqref{g2Expression} implies that the metric $\check g_{m \bar n}$ introduced in \eqref{gexp} is 
still K\"ahler and that the internal part of the eleven-dimensional metric \eqref{Metric11d}
is conformally K\"ahler.

The background also includes a flux for the four-form given by
\ba
 \wh G_{m \bar n r \bar s} & = \eppr G_{m \bar n r \bar s} + \cO(\eppr^3) \, , &
 \wh G_{m n r s} & = \eppr G_{m n r s} + \cO(\eppr^3) \, , \nn 
\ea
\vspace{-1cm}
\ba \label{FluxAnsatz}
 \wh G_{\m\n\r m} &= \e_{\m \n \r} \pa_m e^{ -3 \eppr^2 W} + \cO(\eppr^3)  \,  .
\ea
In order that the eleven-dimensional field equations are solved to order $\alpha^2$ by 
this background the flux $G $ must be 
self-dual in the lowest-order metric $g_{m \bar n}$. This condition allows 
$(2,2)$ and $(4,0)+(0,4)$ components of the flux with respect to the lowest order complex structure. 
The profile of the warp-factor $W$ depends both on the background flux $G$
and the higher-curvature terms through the equation
\ba \label{warpfactoreq}
  & d^\dagger d e^{ 3 \alpha^2 W} *_81  -  \alpha^2 Q_8 + \cO(\alpha^3) = 0\ ,
\ea
where 
\beq \label{def-Q8}
   Q_8 = -  \fr12 G \we G - 3^2 2^{13} \alpha^2 X_{8} 
         =  -  \fr12  G \we G  + 3072\, c_4 \ . 
\eeq
In this expression $c_4$
 is the fourth Chern-form evaluated in the metric $g_{m\bar n}$ given by 
 \ba 
c_4 = \fr{1}{8} ( \cR_{m}{}^n \cR_{n}{}^m \cR_{r}{}^s \cR_{s}{}^r - 2 \cR_{m}{}^n \cR_{n}{}^r \cR_{r}{}^s \cR_{s}{}^m) \ .
\ea
Asserting that $Y_4$ is compact, the warp-factor equation \eqref{warpfactoreq} implies the global consistency condition
\beq \label{tadpole}
   \frac{1}{3^2 2^{14}} \int_{Y_4} G \we G =  \frac{\chi(Y_4)}{24} \ ,
\eeq
where $\chi(Y_4) = - 4! \int_{Y_4} X_8= \int_{Y_4} c_4 $ is the Euler number of $Y_4$.
This implies that, by using the self-duality of $G$, 
 the higher-derivative terms cannot be consistently ignored if one 
allows for a background flux.\footnote{The  numerical 
factor in \eqref{tadpole} can be attributed to our normalisation of $G$
with $\alpha$ and can be removed when moving to quantised 
fluxes $G^{\rm flux} = \frac{1}{3\, 2^6 \sqrt2 } G$.}

Supersymmetry of the solution \eqref{Metric11d}, \eqref{FluxAnsatz} has not been demonstrated 
to order $\alpha^2$. This can be traced back to the fact that the supersymmetry variations
of the fermions have not been derived to this order. In \cite{Grimm:2014xva} a proposal 
was made for the gravitino variations including order $\alpha^2$ terms based on \cite{Lu:2003ze,Lu:2004ng},
and supersymmetry was successfully checked. Asserting that 
the gravitino variations, as the ones proposed in \cite{Grimm:2014xva}, 
are unchanged at linear order in  $\alpha$, then the flux $G$ satisfies 
\beq \label{G40}
   G_{mnrs} = 0\ ,
\eeq
i.e.~its $(4,0)$ component vanishes, and respects the primitivity condition 
\beq \label{primitivitycond}
 G \we J = 0\ .
\eeq
In this work we will provide further evidence that the solution preserves 
supersymmetry. We derive the three-dimensional action and demonstrate 
compatibility with three-dimensional $\cN=2$ supersymmetry. Furthermore, we show 
that the scalar potential vanishes to order $\alpha^2$ 
when imposing \eqref{G40} and \eqref{primitivitycond} .

\subsection{Considered variations of the background solution}

Having reviewed the background solution in subsection \ref{background_solution}, 
we now include a well-defined set of variations around 
this vacuum and recall the derivation of their effective action. 

Firstly, we will include vectors $A^i$ that arise in perturbations of the M-theory three-form 
$\wh C$. These correspond to extra terms
in the expansion of $\wh G$ of the form 
\beq \label{deltaG}
   \d \wh G = F^i \we \o_i^\tb{v} \,, 
\eeq
where $F^i = d A^i$ are the field strengths of $A^i$,
and $\o_i^\tb{v}$ are two-forms on the internal manifold. 
Importantly, it was argued in \cite{Grimm:2014efa} that in 
the expansion $\o_i^\tb{v} = \o_i^\tbzero{}^\tb{v} + \eppr^2  \o_i^\tbtwo{}^\tb{v}$
only the harmonic 
part of $\o_i^\tb{v}$ contributes in the effective action. We 
may pick $\o_i^\tbzero{}^\tb{v}$ to be harmonic and drop $ \o_i^\tbtwo{}^\tb{v}$.
This implies that $\o_i^\tb{v}$ can be chosen to be harmonic 
$(1,1)$-forms and one has $i=1,\ldots,\text{dim}(H^{1,1}(Y_4))$,
where $H^{1,1}(Y_4)$ is the $(1,1)$-form cohomology of $Y_4$ whose  dimension 
is independent of the metric chosen on $Y_4$.

Secondly, one can analyse the K\"ahler structure deformations
of the conformally K\"ahler metric in \eqref{Metric11d}. We introduce variations 
\beq \label{Kaehler_fluc}
\d g_{m \bar n} = i \d v^i \, \o_{i \, m \bar n }^\tb{s}\ ,
\eeq
where $ g_{m \bar n}$ is the K\"ahler metric given in \eqref{gexp}. The $\d v^i$ correspond 
to scalars in the three-dimensional effective theory, while the $ \o_{i \, m \bar n }^\tb{s}$
is a set of two-forms on $Y_4$ chosen to ensure that the K\"ahler condition remains to 
be satisfied. Remarkably, expanding $\o_i^\tb{s} = \o_i^\tbzero{}^\tb{s} + \eppr^2  \o_i^\tbtwo{}^\tb{s}$
it was again shown in \cite{Grimm:2014efa} that only the harmonic part of $\o_i^\tb{s} $
contributes in the effective action. We therefore drop $ \o_i^\tbtwo{}^\tb{s}$
and chose $\o_i^\tbzero{}^\tb{s}$ to be the same harmonic $(1,1)$-forms 
as in \eqref{deltaG} with $i=1,\ldots, \text{dim}(H^{1,1}(Y_4))$, i.e.~we 
set 
\beq
   \o_i^\tbzero{}^\tb{s} = \o_i^\tbzero{}^\tb{v} = \o_i \ ,
\eeq
where $\o_i$ are the harmonic $(1,1)$-forms in the Ricci-flat zeroth-order metric $g_{m\bar n}$.
In the following it turns out to be convenient to define scalars $v^i$ containing the background value 
of $g_{m \bar n}$ by setting 
\beq \label{flucg}
  g_{m \bar n} +  \d g_{m \bar n} = i v^i \o_{i \, m \bar n }
\eeq

When discussing K\"ahler structure deformations one has to carefully 
vary the complete background solution. In particular, all 
 metric dependent quantities, such as the scalar 
function $Z$ introduced in \eqref{def-Z}, vary non-trivially. The 
second order corrections to the background turn out to be crucial when 
determining the mass of the fields $\d v^i$. Recall that in general the primitivity condition 
\eqref{primitivitycond} is not preserved by all $\d v^i$. 
This implies that one expects a scalar potential depending on the 
flux $G$ as studied in the Calabi-Yau fourfold reductions with 
fluxes in \cite{Haack:1999zv,Haack:2001jz}. It could, moreover, be the 
case that the higher-curvature terms induce additional potential terms. 
We will show in the next subsection that this is not the 
case when including $\alpha^2$ corrections both in 
the background solution and in the eleven-dimensional action.

\subsection{Scalar potential}

In this subsection we discuss the derivation of the scalar potential 
for the K\"ahler structure fluctuation $\delta v^i$ introduced in \eqref{Kaehler_fluc}.
As already pointed out, we expect a flux-induced scalar potential for 
all fluctuations that do not respect the primitivity condition \eqref{primitivitycond}. 

To begin with we consider the terms containing $\hat C$ without derivative. 
Considering the pure three-dimensional space-time part for $\hat C$ one 
easily sees
\beq \label{CSvanish}
 - 
 \int \big( \fr16  \wh C \we \wh G \we \wh G + 3^2 2^{13}    \wh C \we \wh X_8\big) \Big|_{\rm pot}= 0 \ ,
\eeq
which can be traced back to the fact that this combination is proportional to the tadpole constraint \eqref{tadpole}.
A pure flux-induced potential term arises from the reduction
\beq
- 
    \int  \fr12 \wh G \we \wh * \wh G \Big|_{\rm pot} = -  \eppr^2 
    \int_{\cM_3}   \ast_3 1 \int_{Y_4}  \frac{1}{2} G \wedge \ast' G  \ ,
\eeq
where $ \ast'$ is the Hodge star of the perturbed internal metric \eqref{flucg}. 
In order to derive the full flux-induced potential, however, we need to also 
dimensionally reduce the higher-curvature terms. Inserting the fluctuated ansatz 
into the $\wh R^4$-corrections to the eleven-dimensional action we find 
\ba \label{higher-curvature-potential}
   \int \wh t_8 \wh t_8 \wh R^4 \wh * 1& =   
     \int_{\cM_3}   \ast_3 1 \int_{Y_4} \Big(1536 \, c_4  - 768 \,  \d v^i \d v^j   (  \nabla_a \nabla^{a} Z ) \, \omega_{i m \bar n} \omega_{i}{}^{\bar n m} \ast 1 \Big)    \nn \\
- 
   \int  \fr1{24} \wh \e_{11} \wh \e_{11}  \wh R^4 \wh * 1 & =  
     \int_{\cM_3}   \ast_3 1 \int_{Y_4} 1536\, c_4 \; .
\ea
We thus encounter the integral over the forth Chern-form $\int_{Y_4} c_4 = \chi(Y_4)$ 
and \eqref{tadpole} can be used to replace these terms with a flux-dependent contribution 
proportional to $\int_{Y_4} G \wedge  G$. 
Furthermore, there appears to be an additional mass term for the fluctuations $\delta v^i$ involving 
the higher-curvature invariant $Z$. However, we still need to dimensional reduce the zeroth order 
action inserting the $\eppr^2$-corrected background solution. Performing this reduction one finds
\beq \label{R-alpha-pot}
  \int \wh R \wh * 1  =  \eppr^2 
   \int_{\cM_3} \ast_3 1 \int_{Y_4} 768 \,  \d v^i \d v^j   (  \nabla_a \nabla^{ a} Z ) \, \omega_{i m \bar n} \omega_{i}{}^{\bar n m} \ast 1\ ,
\eeq
which precisely cancels the $Z$-dependent mass-term arising from the higher-curvature reduction in \eqref{higher-curvature-potential}.

In summary, adding all terms \eqref{CSvanish}-\eqref{R-alpha-pot} one finds the 
scalar potential term
\ba
S_{\rm pot}= -  \fr{\eppr^2}{4 \k_{11}^2} \int_{\cM_3} \ast_3 1  \int_{Y_4}  \frac{1}{2} \Big( G \wedge \ast' G - G \wedge  G \Big)\ .
\ea
This term has to be still Weyl-rescaled to bring the action into the three-dimensional Einstein frame. The rescaled result 
will be given in \eqref{flux-induced-terms}. As expected one can check that the scalar potential 
vanishes for primitive $(2,2)$-fluxes, i.e.~for all $(2,2)$-fluxes satisfying $ G_{m\bar n \r \bar s} J'^{\bar r s} = 0$.
This condition generically fixes a number of deformations $\delta v^i$ in the vacuum. Note that 
this is the only effect stabilising moduli at order $\eppr^2$ in our setting.  


\subsection{Three-dimensional effective action}

Having discussed the scalar potential, we now recall the complete 
three-dimensional effective action for the fluctuations $\delta v^i$ and 
vectors $A^i$ following \cite{Grimm:2014efa}. It was shown in this work that 
it takes the remarkably simple form 
\beq
    \label{FinalResult}
    \k^2_{11} S_{\rm eff} = S_\text{kin} + S_{\rm CS}+S_\text{pot} \ ,
\eeq
with kinetic terms given by 
\ba
 S_\text{kin}  =&  \int_{\cM_3} \Big(  \frac12 R *1 -  \frac12 (G^T_{ij} +\cV_T^{-2} K_i^T  K_j^T ) D v^i \we * D v^j  
    -  \frac12  \cV^{2}_T G_{ij}^T F^i \we * F^j \Big)\, ,
\ea
and flux-induced Chern-Simons terms and scalar potential given by
\ba \label{flux-induced-terms}
    S_{\rm CS}=  \int_{\cM_3}  \frac12  \Theta_{ij} A^i \wedge F^i \ ,\qquad S_\text{pot}  = -\eppr^2  \int_{\cM_3} \ast_3 1  \int_{Y_4}  \frac{1}{8 \cV_0^{3}}\Big( G \wedge \ast' G - G \wedge  G \Big)\ .  
\ea
The Chern-Simons terms are dependent on the fluxes via
$\Theta_{ij} = \frac{\alpha}{2}\int_{Y_4} G \wedge \omega_i \wedge \omega_j$.
In the following we introduce the remaining coefficient functions appearing in \eqref{FinalResult}.

To begin with,  as in the leading order reduction, we use the quadruple intersection numbers 
$K_{ijkl} = \int_{Y_4} \omega_i \wedge  \omega_j \wedge \omega_k \wedge \omega_l$ to define 
\beq \label{def-Ki}
\cV = \frac{1}{4!} K_{ijkl} v^i v^j v^k v^l    \, ,\quad  K_i = \frac{1}{3!} K_{ijkl} v^j v^k v^l \, , \quad K_{ij} = \frac{1}{2} K_{ijkl}  v^k v^l \, \quad K_{ijk} = K_{ijkl}  v^l \ .
\eeq
Here we have used the fully fluctuated $v^i$ defined in \eqref{flucg}. In the background 
they take the value $v^i_0$. In this case we write, for example, $\cV$ as $\cV_0$ and $K_i$ as $K_i^0$. 
$\cV_0$ is simply the background zeroth-order volume of $Y_4$ also given by $\cV_0 = \int_{Y_4} * 1$.

Starting with the classical couplings, one  can next include the warp-factor. It turns out to be 
convenient to define the warped volume and the warped metric as
\beq \label{def-cVW}
\cV_W  = \int_{Y_4} e^{3 \eppr^2 W} * 1\ , \qquad 
G^W_{i j}  = \frac{1}{2 \cV_W} \int_{Y_4} e^{3 \alpha^2 W} \o_{i} \wedge * \o_{j} \ .
\eeq
The dimensionally reduced action also contains the 
first derivatives of the warp-factor with respect to the moduli $v^i$.
They appear only through a covariant derivative 
\beq
\label{Dv1}
D v^i  = d v^i + \alpha^2\frac{1}{\cV_0} d v^j \, v^i_0 \int_{Y_4} \pa_{v^j} W  * 1  \ .
\eeq
The significance of this fact will be discussed in detail in subsection \ref{scaling+warping}, 
where we will recall the invariance of the action under a moduli-dependent scaling symmetry. 
Finally, in order to present to full result \eqref{FinalResult} we have to include the corrections 
due to higher-curvature terms. We first define 
\ba \label{def-Zmmnn}
Z_{m \bar m n \bar n} = \fr{1}{4!} \e_{ m \bar m m_1 \bar m_1 m_2 \bar m_2 m_3 \bar m_3}  \e_{ n \bar n n_1 \bar n_1 n_2 \bar n_2 n_3 \bar n_3} R{}^{\bar m_1 m_1 \bar n_1 n_1} R{}^{\bar m_2 m_2 \bar n_2 n_2} R{}^{\bar m_3 m_3 \bar n_3 n_3}\ .
\ea
This tensor satisfies the identities 
\ba
Z_{m \bar m n \bar n} &= Z_{n \bar m m \bar n} = Z_{m \bar n n \bar m}\ , &
\na^{m} Z_{m \bar m n \bar n} &= \na^{\bar m} Z_{m \bar m n \bar n}  = 0\ ,
\ea
and is related to the third and fourth Chern-form $c_3$, $c_4$ as
\ba \label{def-Z_plain}
&Z_{m \bar m} = i 2 Z_{m \bar m n}{}^n =  \fr12 (* c_3 )_{m \bar m}\, , &
&Z = i 2 Z_{m}{}^m  = *( J \we c_3 )\, , \nn  \\
&* (c_3 \wedge \o_i )  = - 2 Z_{m \bar n}\o_i{}^{\bar n m}\, , &
&Z_{m \bar m n \bar n} R^{\bar m m \bar n n} = *  c_4\ , 
\ea
with $Z$ already given in \eqref{def-Z}.

The tensor $Z_{m\bar n r\bar s}$ is the basic building block to display the corrected 
metric $G^T_{i j}$, the coupling $K^T_{i}$, and the volume $\cV_T$. 
Explicitly they take the form
\bea
G^T_{i j} & =& G^W_{i j} + 256 \eppr^2  \fr{1}{\cV_0^2} \int_{Y_4} Z * 1  \int_{Y_4} \o_i{}_{m \bar n } \o_j^{\bar n m} * 1 \nn  \\ 
 && -  256 \eppr^2 \fr{1}{\cV_0} \int_{Y_4} \bls Z \o_i{}_{m \bar n } \o_j^{\bar n m}  + 12 Z_{m \bar n r \bar s} \o_j^{\bar n m}   \o_i^{\bar s r} \brs *1 \nn \,  ,\\
K_i^T & =& K^0_i + \eppr^2 \int_{Y_4}  \bls  \frac{K^0_i}{\cV_0} ( 3  W  - 128 Z )* 1 - 1536  Z_{m \bar n} \o_i^{\bar n m}  * 1 \brs \, ,  \nn \\
\cV_T & = & \cV_W +256  \eppr^2 \int_{Y_4} Z * 1 \, ,
\eea
where all quantities are evaluated in the background $v^i_0$.
The coefficients in the expressions for $G^T_{i j}$,$K^T_{i}$, and $\cV_T$ first appear to 
be unrelated. However, they are in fact precisely taking values so as to ensure the identity 
\ba \label{usefullid}
(G^T_{ij} +\cV_T^{-2} K_i^T  K_j^T )  = G^T_{ab} ( \d_i{}^a - \fr{1}{\cV_0} v_0^a K^0_i ) ( \d_j{}^b - \fr{1}{\cV_0} v^b_0 K_j^0 ) \ ,
\ea
which holds in the background $v^i_0$. 
As we will demonstrate in the next section, this identity is one of the crucial ingredients to ensure 
supersymmetry of the three-dimensional effective action.  

\subsection{Warp-factor scaling symmetry and integration} \label{scaling+warping}

In this subsection we will have a closer look at the couplings in \eqref{FinalResult}. In particular,
it was observed in \cite{Grimm:2014efa} that the three-dimensional effective action 
permits a scaling symmetry involving the rescaling of the warp-factor. 
More precisely, the action turns out to be invariant under the symmetry
\ba 
\label{symmetry}
&W \ra W + \Lambda  \ , & 
&   v^i \ra e^{ - \eppr^2 \Lambda} v^i \ .
\ea
for any scalar function $\Lambda = \Lambda(v^i)$
that can be space-time dependent. The scalars $v^i$ in \eqref{FinalResult} therefore 
have to appear with a covariant derivative \eqref{Dv1}, which can be extended to include the fluctuations by writing 
\ba\label{Dv2}
D v^i  &= d v^i + \alpha^2 \cW_j d v^j \, v^i \ , &
 \cW_j &= \frac{1}{\cV}  \int_{Y_4} \pa_{j} W  * 1\ .
\ea
It is conceivable that this scaling invariance persists beyond the $\alpha$-order testable 
in the current reduction. It is also interesting to note that one 
can introduce a potential $\cW$ for the connection in \eqref{Dv2} as 
\ba \label{potentialcW}
   \cW_j  &=\pa_{j} \Big( \frac{\cW}{\cV} \Big)\ , &
    \cW(v^i) &=   \frac{1}{4! }  \int_{Y_4} W  J^4 \ ,     
\ea
where $J = v^i \omega_i$ contains the fluctuated K\"ahler moduli. 

The scaling symmetry fixes a number of the warp-factor dependent terms in \eqref{FinalResult}
and one readily infers a potential $\cW$ that appears in these couplings.
However, there is one contribution proportional to 
$\int_{Y_4} W \omega_i \wedge \omega_j \wedge J \wedge J$ that appears to be special. 
It arises by expanding 
\beq \label{expandGW}
G^W_{i j}  = - \fr{1}{2 \cV} K_{ij} + \fr{1}{2 \cV^2}  K_i K_j - \frac{3}{4 \cV} \int_{Y_4} W \o_{i} \wedge \o_{j} \wedge J \wedge J + \fr{3}{2 \cV^2}  K_{i j} \int W *1  \ ,
\eeq
where we have used\,\footnote{Note that this relation only holds for harmonic forms $\omega_i$.}
\beq
   *\o_i =-\fr12 \omega_i \wedge J \wedge J + \fr{1}{6 \cV } K_i J^3\ .
\eeq
At first, one might have suspected that all terms in \eqref{expandGW} 
arise as derivatives of $\cW$ as well. 
However, evaluating\,\footnote{A simple
way to show the first identity is to split the integral $\int W \omega_i \wedge J^3 \propto \cW K_i$, by using that $\omega_i \wedge J^3$ is 
harmonic.}
\ba
   \partial_j \int W \omega_i \wedge J^3 &=  3! K_i  \cW_j + 3! \frac{\cW}{\cV}K_{ij}\ , &
   \partial_j \cW_i &= \frac{1}{4! \cV} \int_{Y_4} (\partial_i \partial_j W) J^4 \ , 
\ea
one infers that there is no term proportional to $\int W \o_i \wedge \o_j \wedge J^2$. 
This is a first example of a situation where one can connect couplings with zero and 
one index, but new structures arise at the two-index level. 
We discuss similar issues arising in the higher-derivative sector next. 

In order to integrate terms in the higher-derivative sector, one might want to start with 
the scalar function
\beq \label{def-cZ}
   \cZ(v^i) = \frac{1}{4!} \int_{Y_4} Z J^4 = \int_{Y_4} J \wedge c_3\ .
\eeq
where we have used \eqref{def-Z_plain} and view $\cZ$ as a function of the fluctuated 
moduli $v^i$. It is then straightforward to derive 
\beq
  \cZ_i = \partial_i \cZ = \int_{Y_4} \omega_i \wedge c_3 = - 2  \int_{Y_4} Z_{m \bar n}\o_i{}^{\bar n m} *1\ ,
\eeq
where we again inserted \eqref{def-Z_plain}. Note that when written with the Chern-form $c_3$ it 
is obvious that $\cZ_i$ is actually constant such that $\partial_j \cZ_i = 0$. 
Thus, in complete analogy to the warping terms, there appears to be no obvious 
potential that admits the two-index terms
\ba \label{two-index-Zterms}
  &\int_{Y_4} Z_{m \bar n r \bar s} \o_j^{\bar n m}   \o_i^{\bar s r} *1\ , &
 & \int_{Y_4} Z \omega_i \wedge \omega_j \wedge J\wedge J \, ,
\ea
 as derivatives. 
 We will have to address precisely these obstacles when showing the supersymmetry 
of the effective action in  next section. 

To close this section let us point out that the two terms in 
 \eqref{two-index-Zterms} are just part of a set of higher derivative 
 terms of the form 
 \beq
   \cX^{(r)}_{ijkl} = \int_{Y_4} \omega_i \wedge \cR_{m_1 \bar m_1} \wedge \cR_{m_2 \bar m_2} \wedge \cR_{m_2 \bar m_2}\, \omega_i^{\bar n_1 n_1} \omega_k^{\bar n_2 n_2} \omega_l^{\bar n_3 n_3} \ (\mathcal{Y}_{(r)})^{m_1 \bar m_1 m_2 \bar m_2 m_3 \bar m_3 }_{n_1 \bar n_1 n_2 \bar n_2 n_3 \bar n_3}\ ,
 \eeq
where the $\mathcal{Y}_{(r)}$ are defined to encode all possible index contractions of $m_p$ with $n_q$.
The two terms in \eqref{two-index-Zterms} arise when contracting a 
particular set of $\cX^{(r)}_{ijkl} $ with $v^k$ and $v^l$. It would be very interesting to 
study the properties of such $\cX^{(r)}_{ijkl}$. In particular, the variation of these terms 
with the moduli $v^i$ might uncover interesting relations. Furthermore, it is worth stressing 
that the terms $\cX^{(r)}_{ijkl}$ including the contractions \eqref{two-index-Zterms}
depend on the chosen forms $\omega_i$, i.e.~not just on the class of $\omega_i$, 
for all appearing two-forms. In our study the $\omega_i$ were always harmonic, but it  
would be interesting to check if there are linear combinations of the $\cX^{(r)}_{ijkl}$ or 
its $v^p$ contractions that only depend on the cohomology class of the two-forms.  

\section{Demonstrating the supersymmetric structure} \label{susy_structure}

In this section we determine the K\"ahler potential and complex coordinates compatible 
with $\cN=2$ supersymmetry in three dimensions. Our starting point will be the 
three-dimensional effective action \eqref{FinalResult} obtained by 
dimensional reduction. We discuss its supersymmetric structure both in the frame when working 
with vectors $A^i$ and in the dual frame when the vectors are replaced by scalars $\rho_i$.

\subsection{Comparing the reduction result with $\cN=2$ supergravity}

It turns out to be convenient to first work with three-dimensional vector multiplets
with bosonic fields $(L^i,A^i)$ and only later switch to chiral multiplets with complex scalars 
$T_i$. The kinetic terms of an ungauged $\cN=2$ supergravity theory can be written as
\beq \label{GenN=2withL}
S_\text{kin}  =  \int_{\cM_3} \Big(  \frac12 R *1 +\frac{1}{4}\tilde K_{L^i  L^j} d L^i \we * dL^j  +  \frac{1}{4} \tilde K_{L^i  L^j} F^i \we * F^j \Big)\, .
\eeq
In this expression $\tilde K_{L^i L^j}$ can be determined from a so-called 
kinetic potential $\tilde K(L)$ via $\tilde K_{L^i L^j} = \partial_{L^i} \partial_{L^j} \tilde K$.
Dualising the vector $A^i$ in the vector multiplet one can translate the three-dimensional theory 
into an action for complex scalars $T_i$ with kinetic terms given by a K\"ahler potential $K(T,\bar T)$. 
The action then takes the form 
\beq
 \label{GenN=2withLT}
S_\text{kin}  = \int_{\cM_3} \Big(  \frac12 R *1 - K_{T_i \bar T_j} d T_i \we * d\bar T_j\Big) \, ,
\eeq
where $K_{T_i \bar T_j} = \partial_{T_i} \partial_{\bar T_j} K$ is the K\"ahler metric. 
Note that $\R T_i$, $K$ and $L^i$, $\tilde K$ are related by a Legendre transform as
\ba \label{def-TfromtildeK}
  T_i &= \tilde K_{L^i} + i \rho_i \ , &
   K &= \tilde K - \frac{1}{2} (T_i + \bar T_i) L^i\ ,
\ea 
where $\rho_i$ is the three-dimensional scalar dual to the vector $A^i$. 
One can now straightforwardly derive that 
$K_{T_i \bar T_j} = - \frac{1}{4} \tilde K^{L^i L^j}$, which uses 
the inverse of $\tilde K_{L^i L^j}$. Note that $K$ is independent of the scalar $\rho_i$
and thus a function $K(\R T_i)$. It is useful to recall the inverse transformation
\beq \label{LasKT}
   L^i  = - 2 K_{T_i} \ ,
\eeq
where $K_{T_i} = \partial_{T_i} K$.
The theory formulated in the $T_i$ coordinates can admit 
a scalar potential of the form  
\beq \label{gen_N=2scalarpot}
    S_{\rm pot} = - \int_{\cM_3} \big(K^{T_i \bar T_j} \partial_{T_i} \cT \partial_{\bar T_j} \cT - \cT^2 \big) *1 + e^{K} \big(K^{T_i \bar T_j}    
    D_{T_i} W \overline{D_{T_j} W}  - 4 |W^2|) *1 \ , 
\eeq
where $K^{T_i \bar T_j} $ is the inverse of the K\"ahler metric $K_{T_i \bar T_j} $. 
Here $\cT$ is a real function of the fields $T_i$, while $W$ is a holomorphic superpotential in 
the $T_i$. The latter transforms non-trivially under K\"ahler transformations and 
therefore appears with the K\"ahler covariant derivative $D_{T_i} W = \partial_{T_i} W + K_{T_i} W$.

To read off $\tilde K_{L^i L^j}$ we compare 
 the action \eqref{GenN=2withL} with the result from the dimensional reduction \eqref{FinalResult}.
 We first read off the coefficient of the $F^i \wedge * F^j$ term and identify
\bea \label{id_tildeK}
\tilde K_{L^i L^j} \big|= - \tfrac12 \cV^{2}_T G_{ij}^T \ .
\eea
Here we have used the notation $f(v^i)| = f(v^i_0)$, i.e.~the vertical dash denotes evaluation in the background setting 
all fluctuations $\delta v^i = 0$.
Supersymmetry implies that for the correct definition of $L^i$, this metric has 
to match the one in front of $dL^i \wedge * dL^j$. 
Applied to \eqref{FinalResult} this implies the relation  
\ba
  \cV^{2}_T G_{ij}^T = (G^T_{cd} +\cV_T^{-2} K_c^T  K_d^T )  ( \d_{a}^c + v^c_0  \cW^0_a) ( \d_{c}^d +  v^d_0  \cW^0_b ) \fr{\pa v^a}{\pa L^i} \Big|\, \fr{\pa v^b}{\pa L^j}  \Big|\ ,
\ea
where $\cW_a^0 = \cW_a|$ is defined in \eqref{Dv2} and evaluating in the background.
Then using \eqref{usefullid} we find that 
\ba \label{dLbydvfromReduction}
\pa_j L^i |  \equiv \fr{\pa L^i}{ \pa v^j} \Big| & =  \fr{1}{\cV_T} \Big( \d_k^i - \fr{v^i_0}{\cV_0}  K^0_k \Big) \big( \d_{j}^k + v_0^k \cW^0_a \big) \ ,
\ea
where as above we abbreviate derivatives with respect to $v^i$ as $\partial_i \equiv \frac{\partial}{\partial v^i}$.
It turns out to be complicated to integrate this condition. This can be traced back to the fact that 
there is an evaluation and, as we discuss below, the fundamental objects to define $L^i$ itself 
might be more involved. Nevertheless, we can already make some interesting observations.
Firstly, the higher-curvature corrections only appear through $\cV_T$ in \eqref{dLbydvfromReduction}. 
One suspects that this can only be true in the background. In fact, we might imagine 
that  $\pa_j L^i$ contains a term
\beq
  \pa_j  L^i  \supset  v^i \int_{Y_4} \big[ Z_{m \bar n } \o_j^{\bar n m}  -  2 Z_{m \bar n r \bar s} \o_j^{\bar n m}   \o_k^{\bar s r} v^k \big] *^\tbzero 1\ ,
  \eeq
which trivially gives zero when evaluated at $v^i_0$. Terms of this type, however, will turn out to be crucial in 
order to determine the underlying objects of the theory. In contrast, artificially switching off the higher-curvature 
corrections in \eqref{dLbydvfromReduction} one finds that the $L^i$ in the presence of warping actually 
takes the simple form 
\beq
   L^i = \frac{v^i}{\cV_W}\ ,
\eeq
where $\cV_W$ is the warped volume \eqref{def-cVW} now evaluated as a function of the perturbed $v^i$.

As a second requirement of supersymmetry we note that \eqref{def-TfromtildeK} implies 
\beq
 \pa_i \R T_j\big| =   \tilde K_{L^j L^k} \pa_i L^k  \big|\ .
\eeq
Using \eqref{id_tildeK} and \eqref{dLbydvfromReduction} we conclude that
\bea
\pa_j \R T_i  \big|
 &=&  K_{ij}^0 + 3  \eppr^2 K^0_i   \cW_j^0 +  \fr32 \eppr^2 \int_{Y_4} W \o_i \we \o_j \we J \we J \, \big|  
  \\ 
 &&   -   256 \eppr^2 \fr{1}{\cV_0} K^0_{i j } \cZ_0 -
  1536 \eppr^2 \fr{1}{\cV_0} K^0_j  \cZ_i^0
  \nn \\
 &&+ 256 \eppr^2  \int_{Y_4} Z \o_i \we \o_j \we J \we J \, \big|   + 6144 \eppr^2 \int_{Y_4} \o_i{}^{\bar n m} \o_j{}^{\bar s r}Z_{m \bar n r \bar s } *1\,\big| \ ,\nn
 \label{dTbydvfromReduction}
\eea
where  
$K_{ij}^0$ and $K_i^0$ are introduced in \eqref{def-Ki} and evaluated at $v_0^i$. 


\subsection{K\"ahler potential and coordinates as a $\d v$ expansion}

In the previous section we have deduced the expressions for $\pa L^i/ \pa v^j$ 
and $\pa \R T_j /\pa v^i$ when evaluated in the background $v^i = v_0^i$.
We will next try to infer directly the coordinates $T_i $ and the K\"ahler potential 
$K$. In order to do this we view $T_i$ and $K$ as being given by 
an expansion both in $\eppr$ and $\delta v^i$ by writing  
\ba
\R T_i & = \R T_i^\tbzero + \eppr^2  \R T^\tbtwo_i \, ,& 
\R T^\tbtwo_i & = \R T^\tbtwo_i |  + \pa_{j} \R T^\tbtwo_i | \d v^j  + \fr12 \pa_j \pa_k \R T^\tbtwo_i |  \d v^j \d v^k \, ,\nn \\
K & = K^\tbzero + \eppr^2 K^\tbtwo \, ,& 
K^\tbtwo & =  K^\tbtwo |  + \pa_j K^\tbtwo | \d v^j  + \fr12 \pa_j \pa_k K^\tbtwo |  \d v^j \d v^k \ .
\ea
In the following we derive as much information as possible about the coupling functions that appear in this expansion by comparing to the reduction result. 

As a first step, recall that the zeroth order result in $\alpha$ was already determined in 
\cite{Haack:1999zv,Haack:2001jz}. With our above expressions one can check that 
\beq\label{KTclassical}
   K^\tbzero =  - 3 \log(\cV)\ , \qquad \R T^\tbzero_i = K_i\ ,
\eeq 
where now $\cV$ and $K_i$ depend on the varying $v^i$. 
At the next order in $\alpha$ we note that there are only few objects with zero or
 one index $i$ that are non-trivial 
in the background. More precisely, one can write
\beq \label{Ktwo-back}
         K^\tbtwo | = \frac{\mu_1}{\cV_0}\cZ_0+ \frac{\mu_2}{\cV_0}\cW_0\ ,
\eeq
where $\cZ$ and $\cW$ are defined in \eqref{def-cZ} and \eqref{potentialcW}. 
The constants $\mu_1,\mu_2$ are undetermined  at this point. Clearly, the constant 
shifts in $K$ are unimportant for the derivation of the K\"ahler metric. However, the form 
of \eqref{Ktwo-back} might 
hint towards the fully moduli-dependent form of $K$. To fix the coefficients 
$\mu_2$ one might be inclined to use the scaling symmetry \eqref{symmetry}.
Together with the classical form of $K$ one then infers that an invariant 
$K$ requires $\mu_2 = -12$.

We 
can proceed similarly for the one-index quantities. We first make an ansatz using all 
one-index building blocks we have encountered so far by setting
\ba \label{TK-Ansatz}
  \R T^\tbtwo_i | &=  \tilde \nu_1 \cZ_i  +
   \tilde \nu_2  \cV_0 \cW_i^0 
  +   \tilde \nu_3 K^0_i  \cZ_0 +\tilde \nu_4 K^0_ i \cW_0 \ ,  \nn \\
  \partial_{i} K^\tbtwo | &=  \frac{\tilde \mu_1}{\cV_0} \cZ_i +
   \tilde \mu_2  \cW_i^0
  +  \frac{\tilde \mu_3}{\cV_0} K^0_i \cZ_0 + \frac{\tilde \mu_4}{\cV_0}  K^0_ i \cW_0 \ . 
\ea
The constant coefficients $\tilde \nu_\alpha,\tilde \mu_\alpha$ are not determined 
at this point, since there are no direct relations fixing the background values of $T_i$ and $\partial_i K$.
To fix at least some of the coefficients in \eqref{TK-Ansatz} 
one can again use the symmetry \eqref{symmetry}. Note that $T_i$ are proper complex 
coordinates that should be invariant under \eqref{symmetry}. This suggests 
that $\tilde \nu_4 = 3$ and $ \tilde \nu_2 = 0$, where we have used that the leading 
contribution to $T_i$ is of third power in $v^i$ as in \eqref{KTclassical}.
In contrast, we note that $K$ should be invariant under \eqref{symmetry},
while $\partial_{i} K^\tbtwo $ should transform as a derivative and 
therefore contain the connection $\cW_i$. Using again the leading form \eqref{KTclassical}
and the expression \eqref{potentialcW}
one concludes $\tilde \mu_2 = -12$ and $\tilde \mu_4 = 0$.

In contrast to \eqref{Ktwo-back} and \eqref{TK-Ansatz} the form of $ \pa_j \R T^\tbzero_i |$ and $ \pa_j \R T^\tbtwo_i |$ are fully 
fixed by the reduction and are trivially read off from \eqref{dTbydvfromReduction}
with 
\bea \label{parjT}
\pa_j \R T_i^\tbtwo  \big|
 &=&  3  K^0_i \cW_j^0  + \fr32  \int_{Y_4} W \o_i \we \o_j \we J \we J   \, \big|  
  \\ 
 &&   -   256   \fr{1}{\cV_0} K^0_{i j } \cZ_0 - 1536 \fr{1}{\cV_0} K^0_j  \cZ_i^0
  \nn \\
 && + 256    \int_{Y_4} Z \o_i \we \o_j \we J \we J \, \big|   + 6144   \int_{Y_4} \o_i{}^{\bar n m} \o_j{}^{\bar s r}Z_{m \bar n r \bar s } *1\, \big| \ .\nn
\eea

All other remaining terms in the expansion \eqref{TK-Ansatz} are also not fully determined 
 by our results obtained from the reduction. However, we can 
use \eqref{LasKT} to show that the general relation
\beq \label{LiasKderivative}
   L^i =- 2 \frac{\partial K}{\partial T_i} = - \frac{\partial K}{\partial v^j}  \frac{\partial v^j}{\partial \R T_i} \ ,
\eeq
together with \eqref{TK-Ansatz} gives
\ba \label{Ltwo}
   L^{\tbtwo i} &= 
   -  K^{ i j} \pa_j K^\tbtwo | 
   -   \fr{1}{\cV} v^j K^{  i k}  \pa_k T^\tbtwo_j  |
   +    K_{j l m} K^{  i l} K^{ k m} \pa_k K^\tbtwo |\d v^j
   \nn \\ & 
   -   K^{ i k} \pa_j \pa_k K^\tbtwo |\d v^j 
   -   \fr{1}{\cV} K^{ i k}  \pa_k T^\tbtwo_j  | \d v^j
   +   \fr{1}{\cV^2}  K_j v^l K^{ i k}  \pa_k T^\tbtwo_l  |\d v^j
    \nn \\ & 
   +   \fr{1}{\cV} K_{j m n}   K^{ i m} K^{ l n} v^k  \pa_l T^\tbtwo_k  |\d v^j
   -   \fr{1}{\cV}  K^{ i l} v^k  \pa_j \pa_l T^\tbtwo_k  |\d v^j   + \cO(\d v^2)
    \ . 
\ea
From this it is straightforward to evaluate $\partial_i L^j$ and compare the 
result with \eqref{dLbydvfromReduction} in the background $v^i=v^i_0$. One then infers that 
the coefficients in \eqref{TK-Ansatz} have to satisfy the relation 
\ba \label{additational_constraint}
 \pa_i \pa_j  \Re  T^\tbtwo_k v^k |   &
 - \cV  K_{i j k} K^{k l} \pa_l K^\tbtwo | 
 +  \cV \pa_j \pa_k K^\tbtwo |   
  \nn \\ 
  =  & 9 \frac{1}{\cV_0} K^0_{ij} \cW_0 
  + 18 \cV_0 \cW^0_{(i} K^0_{j)}  
  + 12 \cV_0 K^0_{i j k} K_0^{k l} \cW^0_l- \fr32 \int W \o_i \we \o_j \we J \we J \, \big|
  \nn \\ & 
  - 256 \int Z \o_i \we \o_j \we J \we J \, \big|
    +3072  \fr{1}{\cV_0} K^0_{(i} \cZ^0_{i)}
  -  1536 \fr{1}{\cV_0^2} K^0_i K^0_j  \cZ_0
  \nn \\ & 
 - 6144 \int \o_i{}^{\bar n m} \o_j{}^{\bar s r}Z_{m \bar n r \bar s } *1 \, \big|
 + 1536 K^0_{i j k} K_0^{k l} \cZ_l^0
 \ea
Imposing these conditions then implies that we match the metric \eqref{id_tildeK}. 
Note that this analysis can be carried out independent of any gauge fixing
of the scaling symmetry \eqref{symmetry}. 
Also note that our first-order analysis does neither uniquely fix the 
K\"ahler coordinates nor the K\"ahler metric. This can be traced back to the 
fact that we performed the dimensional reduction only to leading order in 
the fluctuations $\delta v^i$. 

In order to fix the coefficients in \eqref{TK-Ansatz} further, one can try to impose 
conditions that might hold also at the higher-derivative level. For 
example, one may suspect that a no-scale condition holds even when including 
$\alpha$-corrections to the action. In three space-time dimensions 
such a condition reads
\beq \label{noscale}
    K_{T_i} K^{T_i \bar T_j} K_{\bar T_j} = 4\ .
\eeq 
It ensures that in the scalar potential \eqref{gen_N=2scalarpot} the negative $-4|W|^2$ term cancels for a superpotential independent 
of $T_i$. Using \eqref{LasKT} and $K^{T_i \bar T_j} = - 4 \tilde K_{L^i L^j} $ one rewrites \eqref{noscale}
as 
\beq
   L^i  \tilde K_{L^i L^j}  L^j = - 4\ . 
\eeq
In the background this expression can be evaluated by using \eqref{Ltwo} together with 
\eqref{parjT} to yield the condition\footnote{ 
We note also that a similar set constraints $\Re T_i  \Re T_j G^{i j} |= L^i L^j G^{-1}_{i j} | = L^i \Re  T_i |= 4 $ and $\partial_k (L^i \Re  T_i)|=0$ can all be satisfied if we demand 
\eqref{noscaleKconstraint}  as well as
\ba 
\Re T^\tbtwo_i | =  \pa_i K^\tbtwo | - \fr{1}{3} K_{i} \pa_j K^\tbtwo v^j |   + 12 \cW^0_i + 3 K^0_i \cW_0  - 4 \fr{1}{\cV_0} K^0_i \cW^0_j v_0^j + 256 \fr{1}{\cV} K^0_i \cZ_0 \, . 
\ea}
\beq
\label{noscaleKconstraint}
\pa_i  K^\tbtwo v^i |  = 2304 \fr{1}{\cV_0}\cZ_0  - 12  \cW_i^0 v_0^i \ .
\eeq
Keeping in mind that we have few objects with zero or one index, one can use this condition 
as a further motivation to make an ansatz for the K\"ahler potential and match the coefficients. 
This will be considered in the following section.

\subsection{Completing the K\"ahler potential and complex coordinates}

In this final subsection we comment on the completion of the K\"ahler potential 
and complex coordinates as a closed expression in K\"ahler deformations. 
Our goal is to replace the $\delta v^i$-expansion \eqref{TK-Ansatz}
with an appropriate ansatz hinting towards the underlying 
structure of the higher-derivative reduction. It should be stressed 
that we are only  able to fully justify the leading terms. However, 
we will also discover an intriguing interplay between warping effects
and higher-curvature terms.

To begin with, let us propose an ansatz for the K\"ahler potential. 
We have noted in \eqref{Ktwo-back} that there are only few objects without 
indices. Using the quantities introduced in \eqref{potentialcW} and \eqref{def-cZ} we 
suggest
\bea \label{K-Ansatz}
   K &=& -3 \log \bigg( \int_{Y_4} e^{4 \eppr^2  W } * 1 + 256 \m  \eppr^2  \int_{Y_4} Z * 1 \bigg) 
 \\ 
       &=&- 3 \log (\cV + 256 \mu \alpha^2 \cZ + 4 \alpha^2 \cW + \cO(\alpha^4 ))\ , \nn 
\eea
where the functions that appear are now viewed as being dependent on the 
fields $v^i$.
In this expression we fixed the factor in front of $\cW$ by the fact that 
$K$ has to be invariant under the symmetry \eqref{symmetry}. The 
factor in front of the $\cZ$ term is not fixed a priori and we have introduced the constant $\mu$
to capture this freedom. 
Let us stress that it is straightforward to compute  the $v^i$ derivatives of $K$ as defined
in \eqref{K-Ansatz}. In particular, one finds
\beq \label{Ki-Ansatz}
  \partial_{i} K =  - 3 \frac{1}{\cV} K_i+  768 \mu \alpha^2 \frac{1}{\cV} \cZ K_i -  768  \mu  \alpha^2 \frac{1}{\cV} \cZ_i - 12 \alpha^2 \cW_i\ .
\eeq
Clearly, in order to compute the actual K\"ahler metric we also have to 
supplement an ansatz for the complex coordinates $T_i$. The involved form of 
the K\"ahler metric determined from the dimensional reduction \eqref{FinalResult} and
the rather simple form of the K\"ahler potential \eqref{K-Ansatz} as a function of
 the $v^i$ suggests that the $T_i$ have to capture most of the non-trivial information about 
the $\cN=2$ system. 



To get some intuitive information about $T_i$, we note that these coordinates 
are expected to linearise the action of M5-brane instantons on divisors $D_i$. In fact,
as discussed in \cite{Witten:1996bn} a holomorphic superpotential of the schematic form $W \propto e^{-T_i}$ can 
be induced by such instanton effects. This implies that the $T_i$ are expected to be integrals 
over divisors $D_i$. We therefore suggest that they take the form 
\ba \label{PotentialAndCoordinates}
T_i & =  \int_{D_i} \Big( \fr{1}{3!} e^{3 \eppr^2  W} J \we J  \we J  + 1536 \eppr^2 F_6 \Big)  + i \rho_i\ ,
\ea
where $D_i$ are $h^{1,1}(Y_4)$ divisors of $Y_4$ that span the homology $H_2(Y_4,\mathbb{R})$.
The six-form $F_6$ in this expression is a function of degrees of freedom associated with the internal space metric and will be responsible for the more complicated higher derivative structures \eqref{two-index-Zterms}. It is constrained by a relation to the fourth Chern form $c_4$ such that $F_6$ determines the non harmonic part of $c_4$ as 
\beq \label{def-F6}
   c_4 = H c_4 + i \pa \bar \pa F_6\ .
\eeq 
This is in analogy to the quantity $F_4$ introduced for $c_3$ in \eqref{def-F4}.
Note that \eqref{def-F6} leaves the harmonic and exact part of $F_6$ unfixed and 
we will discuss constraints on these pieces in more detail below. The justification 
of the first term in Re$T_i$ is simpler. It captures the warped volume of an M5-brane wrapped 
on $D_i$. In fact, the power of the warp-factor turns out to be appropriate
to ensure invariance under the scaling symmetry \eqref{symmetry}, in accord with the 
expectation that $T_i$ is invariant under this symmetry. Remarkably, this definition of 
the K\"ahler coordinates as $D_i$ integrals will help us to obtain the couplings 
$\int e^{3 \alpha^2 W} J \wedge J \wedge \o_i \wedge \o_j$, which, as we stressed in subsection \ref{scaling+warping},
cannot be obtained as $v^i$-derivatives of the considered $Y_4$-integrals. 
Note that the following discussion of the warping is inspired by \cite{Martucci:2014ska}. Here we will adapt and extend the arguments of \cite{Martucci:2014ska} and include the higher-curvature pieces. Interestingly they turn 
out to complete the analysis in an elegant and non-trivial fashion. 

In order to evaluate the derivatives of $T_i$ with respect to $v^i$ and to make contact with 
the K\"ahler metric found in \eqref{FinalResult}, we have to rewrite the integrals over $D_i$ 
into integrals over $Y_4$. Due to the appearance of the warp-factor and the non-closed form 
$F_6$ in \eqref{PotentialAndCoordinates} this is not straightforward. In particular, one cannot 
simply use Poincar\'e duality and write $T_i$ as an integral over $Y_4$ with inserted $\omega_i$. 
Of course, it is always possible to write $T_i$ as a $Y_4$ integral 
when inserting a delta-current localised on $D_i$, i.e.
\beq \label{T_idelta}
\R T_i  =  \int_{Y_4} \Big( \fr{1}{3!} e^{3 \eppr^2  W} J  \we J  \we J    + 1536 \eppr^2  F_6 \Big)  \we \d_i\ ,
\eeq
where $\d_i$ is the (1,1)-form delta-current that restricts to the divisor $D_i$. 
Appropriately extending the notion of cohomology to include currents \cite{GriffithsHarris,BottTu}, we 
can now ask how much $\delta_i$ differs from the harmonic form $\omega_i$ in the same class.
In fact, any current $\d_i$ is related to the harmonic element of the same class $\o_i$ 
by a doubly exact piece as 
\ba
\d_i = \o_i   + i \pa   \bar \pa   \l_i\ .
\label{delToOm}
\ea
This equation should be viewed as relating currents. Importantly, as we assume 
$D_i$ and hence $\delta_i$ to be $v^i$-independent, the $v^i$ dependence of the harmonic 
form $\o_i$ and the current $\l_i$ has to cancel such that $\partial_j \o_i =  - i\pa   \bar \pa  \pa_j  \l_i$. 
Importantly, once we determine $\partial_j \R T_j$ we can express the result as $Y_4$-integrals without 
invoking currents. We therefore need to understand how each part of $T_i$ 
varies under a change of moduli. This will also fix the numerical factor in front of $F_6$ in \eqref{PotentialAndCoordinates}.

In order to take derivatives of $T_i$ we first use the fact that $D_i$ and hence $\d_i$ 
are independent of the moduli $v^i$, which implies 
\beq \label{eqdelta}
\partial_j \R T_i = \int_{Y_4} \Big( \fr{1}{2} e^{3 \eppr^2  W} \omega_j  \we J  \we J    +  \fr{1}{2} \alpha^2 \partial_j W J  \we J  \we J  +  1536 \eppr^2  \pa_j F_6 \Big)  \we \d_i\ .
\eeq
We next claim that we can replace $\d_i$ with $\omega_i$ such that finally 
\beq \label{eqomega}
\partial_j \R T_i = \fr{1}{2}   \int_{Y_4} e^{3 \eppr^2  W} \omega_i  \we \omega_j \we J  \we J    +  \fr{1}{2} \alpha^2  \int_{Y_4} \partial_j W \o_i \we J  \we J  \we J  +  1536 \eppr^2  \int_{Y_4} \o_i\we  \pa_j F_6  \ .
\eeq
Note that by using \eqref{delToOm} the two expressions \eqref{eqdelta} and \eqref{eqomega} only differ by a
term involving $\pa   \bar \pa   \l_i$. By partial integration this term is proportional to  
\bea \label{extraterm}
&&\int_{Y_4} \l_i \pa \bar \pa \Big( \fr{1}{2} e^{3 \eppr^2  W} \omega_j  \we J  \we J    +  \fr{1}{2} \alpha^2 \partial_j W J  \we J  \we J  +  1536 \eppr^2  \pa_j F_6 \Big)  \nn\\
&=& \int_{Y_4} \l_i  \Big( \fr{1}{2} \pa \bar \pa (e^{3 \eppr^2  W}) \omega_j  \we J  \we J    +  \fr{1}{2} \alpha^2 \pa \bar \pa ( \partial_j W ) J  \we J  \we J  +  1536 \eppr^2  \pa \bar \pa \pa_j F_6 \Big) \ . 
\eea
It is now straightforward to see that the terms multiplying $\l_i$ are simply the $\partial_j$ derivative of the 
warp-factor equation \eqref{warpfactoreq}. One first writes \eqref{warpfactoreq} as
\beq \label{warprewrite}
   d^\dagger d e^{ 3 \alpha^2 W} *_8 1 -  \alpha^2 Q_8  = - \frac{1}{3} i \pa \bar \pa  (e^{ 3 \alpha^2 W} ) \wedge J\wedge J \wedge J - \alpha^2 Q_8\ .
\eeq
Then one takes the $v^j$-derivative of \eqref{warprewrite} by using 
the fact that $Q_8$ is given via \eqref{def-Q8} and \eqref{def-F6}. The moduli dependence of $Q_8$ only arises from the 
term involving $F_6$, i.e.~one has $\pa_i Q_8  = i 3072 \pa \bar \pa \pa_i F_{6} $. Hence one finds exactly 
the terms in  \eqref{extraterm} such that this $\l_i$ dependent part of the $T_i$ variation vanishes due to the warp-factor equation \eqref{warpfactoreq}.

The final expression \eqref{eqomega} is written using \eqref{def-Ki} and \eqref{potentialcW} 
as
\ba
\pa_j \R T_i &   =  \fr12 \int_{Y_4} e^{3 \eppr^2 W} \o_i \we \o_j \we J \we J   
 +  3  \eppr^2 K_i \cW_j 
 +1536  \eppr^2 \int_{Y_4}  \o_i  \we \pa_j F_6   \ .
\label{dTbydvfromAnsatz}
\ea
The $L^i$ coordinates are then computed using \eqref{LiasKderivative} by 
inserting \eqref{Ki-Ansatz} and \eqref{dTbydvfromAnsatz}. This gives the 
result  
\ba
L^i 
  =&  \fr{v^i}{\cV} -  \eppr^2 \fr{v^i}{\cV^2}  (3 \cW + 256 \mu \cZ)  
    + 1536 \eppr^2 \fr{K^{ i j}}{\cV} \Big( \cZ_j -  \int_{Y_4}   J  \we \pa_j F_6 \Big)\ .
\ea
It is then straightforward to derive
\ba
\pa_j L^i =& \fr{\d^i_j}{\cV} 
- \fr{v^i K_j}{\cV^2}
- \fr{\d^i_j}{\cV^2 } (3 \cW + 256 \m \cZ)
-  \fr{1}{\cV }  v^i (3 \cW_j+256 \m   \cZ_j)
+  \fr{1}{\cV^3 } K_j v^i (3 \cW + 512 \m\cZ  ) 
\nn \\ & 
   - \eppr^2 \fr1{ \cV}  768 \m K^{ i m} K^{ k n} K_{ m n  j} \cZ_k - \eppr^2 \fr1{ \cV^2_0} 768 \m K^{ i k} K_j \cZ_k \nn \\& 
+ \eppr^2 \fr1{ \cV}  1536K^{  i m} K^{ k n} K_{ m n  j} \int_{Y_4}  J   \we \pa_k F_6  
+ \eppr^2 \fr1{ \cV^2} 1536 K^{-1 i k} K_j \int_{Y_4}  J   \we \pa_k F_6 
 \nn \\&
- \eppr^2 \fr1{ \cV} 1536 K^{-1 i k} \int_{Y_4}  \o_j  \we \pa_k F_6  
 - \eppr^2 \fr1{ \cV} 1536 K^{-1 i k} \int_{Y_4}  J  \we \pa_j \pa_k F_6  
\label{dLbydvfromAnsatz}
\ea
This allows to determine the derivatives of $F_6$ by 
comparing \eqref{dLbydvfromReduction} and \eqref{dTbydvfromReduction} with \eqref{dLbydvfromAnsatz}  and \eqref{dTbydvfromAnsatz}. We find that
\ba \label{deriv-results}
\int_{Y_4}  \o_i \we \pa_j F_6 |=& 4 \int_{Y_4}  Z_{m \bar n r \bar s} \o_i{}^{\bar n m} \o_j{}^{\bar s r} * 1 
     + \fr{1}{3!} \int_{Y_4}  Z  \o_i {} \we \o_j \we J  \we J  - \fr{K_{i j}}{3! \cV}\cZ - \fr{1}{\cV} K_j\cZ_i 
\nn \\
\int_{Y_4}  J  \we \pa_i \pa_j F_6 |=&  - 4 \int_{Y_4}  Z_{m \bar n r \bar s} \o_i{}^{\bar n m} \o_j{}^{\bar s r} * 1 - \fr{1}{3!} \int_{Y_4}  Z \o_i{} \we \o_j{} \we J  \we J  
\nn \\& 
 - \m \fr{1}{3! \cV}  K_{i j} \cZ - (1-\m) \fr{1}{\cV^2 } K_i K_j \cZ 
+ (2 - \m)\fr{1}{\cV} K_{(i}    \cZ_{j)}
+\frac{1}{2}   (2 - \m)   K_{i j k} K^{k l} \cZ_{l}\ ,
 \ea
in order for the results to match. This implies that the K\"ahler potential \eqref{K-Ansatz} and 
coordinates \eqref{PotentialAndCoordinates} yield the metric matching with the 
reduction result. 

The result \eqref{deriv-results} still depends on the free parameter $\mu$ introduced 
in the K\"ahler potential \eqref{K-Ansatz}. Clearly, one expects that such a freedom is 
not fundamental, but rather due to the fact that we are only able to partially check the result. 
A dimensional reduction including fluctuations to higher order is likely 
fixing $\mu$ unambiguously. Alternatively, we can impose the no-scale condition \eqref{noscale}, 
which we presume persists at higher curvature level. This implies that $\m =1$.  

Let us note that the definition contains two ambiguities. Firstly, we did not specify 
the divisor basis $D_i$ spanning $H_2(Y_4,\mathbb{R})$. This can be shifted by a boundary 
of a seven-chain $\Gamma_i$ without changing the class as
\beq \label{Di-change}
  D_i \ \rightarrow \ D_i + \partial \G_i\ .
\eeq
This would result in a different choice for the currents $\delta_i$ 
and $\lambda_i$ in \eqref{delToOm}. The result 
is a modification of the $\cN=2$ coordinates $T_i$ given in \eqref{T_idelta}.
However, as we have shown above, only the harmonic representative of 
the class enters in the variation $\partial_j T_i$, while $\lambda_i$ drops 
out due to the warp factor equation. In other words, the transformation 
\eqref{Di-change} is actually a symmetry of the K\"ahler metric. 
Secondly, the constraint \eqref{def-F6} is invariant under shifts of $F_6$ by six-forms $\eta_6$, which get annihilated by the derivatives.
In other words, one might transform
\beq
F_6 \ \rightarrow\ F_6 + \eta_6 \ , \qquad \bar \pa \eta_6 = \pa \eta_6 = 0\ .
\eeq
Clearly, this transformation will in general not respect \eqref{deriv-results}. These conditions, however, constrain only 
the harmonic part of $F_6$ and allow for the the symmetry
\beq \label{F6-change}
F_6 \ \rightarrow\ F_6 + d \tilde \eta_4\ .
\eeq
It would be interesting to investigate the implication of the symmetries \eqref{Di-change} and \eqref{F6-change}
in greater detail. This is particularly interesting when including a superpotential explicitly depending on the 
coordinates $T_i$. 

The presence of the $F_6$ term in \eqref{PotentialAndCoordinates} implies, by the above relationship between 
$T_i$ and the action of a probe M5-brane on $D_i$, that higher-derivative corrections are relevant in the M5-brane action. 
Corrections of this type are also required for gravitational anomaly cancellation \cite{Witten:1996hc,Freed:1998tg,Lechner:2001sj} for an M5-brane in the background of eleven-dimensional supergravity. From this anomaly analysis additional metric dependent contributions to the M5-brane action that are related to certain topological classes are expected, in a way similar to the relationship between $F_6$ and $c_4$.  In future work it would be interesting to see if this analysis can be used to infer a more direct definition of the $F_6$ part of the correction in  \eqref{PotentialAndCoordinates} and so prove the constraints  \eqref{deriv-results} that are necessary in our analysis. 

\section{Conclusions}

In this work we continued the study of the three-dimensional effective 
action obtained form dimensionally reducing M-theory on 
eight-dimensional compact manifolds initiated in \cite{Grimm:2014xva,Grimm:2014efa}. The background solutions
contain a warped product of an internal manifold $Y_4$ 
and three-dimensional Minkowski space. The warp-factor 
is induced by non-trivial background fluxes for the 
M-theory four-form field strength, but crucially contains 
contributions from higher-curvature terms of the eleven-dimensional 
action. Global consistency requires these to be included for compact 
internal manifolds. The required higher-curvature terms are suppressed 
by an additional factor of $\alpha^2 \propto \ell_M^6$.
Within an $\alpha$-expansion we were able to consistently include all 
required higher-derivative terms when determining the background 
solution and performing the dimensional reduction.
The resulting three-dimensional action was already presented 
in \cite{Grimm:2014efa}. In this analysis we have so far included the 
deformations of the K\"ahler structure of the geometry and the 
vector modes from the M-theory three-form. Due to the increasing 
computational complexity we performed the derivations only 
to leading non-trivial order in the fluctuations of the K\"ahler structure. 
Nonetheless we were able to identify key features 
of the effective action associated with warping and higher-derivative terms.
One focus of this work was on demonstrating compatibility with the 
structure of a three-dimensional $\cN=2$ supergravity theory. 

As a first result we have shown that the scalar potential 
is only induced by background fluxes. Interestingly the back-reaction 
at order $\alpha^2$ on the background solution was crucial to 
establish this result. Dimensionally reducing the relevant higher-curvature terms \eqref{higher-curvature-potential} 
on a Calabi-Yau fourfold we have found a scalar potential for the K\"ahler 
deformations purely induced by geometry. However, these terms 
cancel precisely with term from the back-reacted metric and led to a confirmation of the flux-induced result of \cite{Haack:2001jz}.
We stress that this cancellation arises only due to the non-trivial eleven-dimensional Weyl rescaling
involving the scalar $Z$ cubic in the Riemann curvature. This rescaling 
can be also performed in eleven dimensions to modify the starting action 
before the dimensional reduction. 

The main focus of this work was the study of the dimensionally reduced 
action with respect to three-dimensional supersymmetry. We used the result of \cite{Grimm:2014efa}
and determined the form of the $\cN=2$ K\"ahler potential $K$ and 
complex coordinates $T_i$. The findings of \cite{Grimm:2014efa} were only at lowest 
order in the fluctuations $\d v^i$, which suggested that we may first determine $K$, $T_i$
as a $\d v^i$-expansion. Already in this evaluation the main complication of 
the dimensional reductions at higher-derivative level became apparent. At
lowest order in $\alpha$ it is straightforward to take the fluctuated 
result for $K$, $T_i$ and `integrate' it into a closed expression depending 
on the K\"ahler form. It is well-known 
that in this case the K\"ahler metric, the K\"ahler potential, and the coordinates $T_i$ 
only depend on topological information, namely the intersection numbers, 
of the manifold $Y_4$. At order $\alpha^2$, however, the result of the 
dimensional reduction contains couplings that are not topological and 
`integrating' these couplings into closed expressions turned out to be challenging. 

As a first example, we found that the three-dimensional action 
contains kinetic terms involving the warp-factor $W$ in the integral 
$\int_{Y_4} W \omega_i \wedge \omega_j \wedge J \wedge J$. 
This integral is not topological and depends on the actual forms 
$\omega_i$ chosen to give its expression. Throughout this 
work $\omega_i$ were the $h^{1,1}(Y_4)$ harmonic representatives 
in the lowest order Ricci-flat metric. We have argued that there is at least no obvious integral over $Y_4$ with only one free-index 
$\omega_i$ that yields the above integral upon taking a $v^i$ derivative.  
Remarkably, at least for the warp-factor terms, one can find a 
way around this problem by defining $T_i$ to be given by  
integrals over divisors $D_i$. Our key observation was that the $v^i$-derivatives 
of the warp-factor equation allows us to write $\partial_j T_i$ as 
$Y_4$-integrals. Furthermore, this $v^j$-variation of $T_i$ was 
argued to only depend on the homology class of the divisor $D_i$ and not 
the precise representative. One might reinterpret this as a statement that 
one now has to consider not only topological integrals, but integrals 
that are `semi-topological' up to usage of the warp-factor equation. 
We believe that a deeper understanding of this fact will shed more light 
onto the proper treatment of effective actions computed in warped string 
compactifications. Importantly, since the warp-factor equation also 
contains higher-curvature terms, we have shown that the 
terms including the warp-factor and the higher-derivative terms cannot be 
analysed independently. 

The analysis of the higher-derivative terms turned out to be even more 
involved. Similar to the warp-factor terms we encountered after 
dimensional reduction non-topological metric-dependent integrals, such 
as $\int_{Y_4} Z \omega_i \wedge \omega_j \wedge J \wedge J$
and $\int_{Y_4} \o_i{}^{\bar n m} \o_j{}^{\bar s r}Z_{m \bar n r \bar s } *1$,
that should arise from a K\"ahler potential. 
As a $\delta v^i$ expansion we have shown compatibility of this 
metric with the existence of a K\"ahler potential and complex coordinates $T_i$.
However, fully integrating  
these expressions to all orders in the fluctuations turned out to be challenging. 
We proposed an expression for $K$ and $T_i$ in \eqref{K-Ansatz} and \eqref{PotentialAndCoordinates}. 
Remarkably, the form \eqref{PotentialAndCoordinates} of $T_i$ is severely constraint by the warp-factor equation. 
It does, however, contain the six-form $F_6$, which is constrained by 
\eqref{def-F6} and therefore contains information about the non-harmonicity of the fourth Chern-form $c_4$ in the 
Ricci-flat metric. The form $F_6$ should capture the 
higher-derivative terms in the three-dimensional action, but we were not able 
to give its full definition including its moduli dependence. The equation \eqref{def-F6}
allows for arbitrary shifts of $F_6$ with harmonic six-forms. Such shifts will in general 
modify $T_i$ and cannot be a symmetry of the system. 
By matching with the result of the dimensional reduction we 
have found that the definition of $F_6$ has to satisfy \eqref{deriv-results}. 
These conditions constrain the harmonic part of $F_6$. It would be of 
crucial importance to give an independent definition of $F_6$ satisfying \eqref{def-F6} 
and \eqref{deriv-results}. 
Our findings suggest already that there is a lot of structure in the 
higher-derivative terms appearing in the effective theory.  

An immediate extension of our analysis is the 
dimensional reduction to next order in the fluctuations 
$\d v^i$, since it would help to further uncover the underlying higher-derivative 
structures. While all four-dimensional couplings at the leading order 
in the $\d v^i$-fluctuations can be written to depend only on the higher-curvature 
quantity $Z_{m\bar m n \bar n}$ a preliminary analysis to the next order suggests
that other higher-curvature couplings are relevant. It would therefore be interesting to 
classify the relevant building blocks in the future.  

Let us close by mentioning a further direction that deserves investigation.
The presented results only deal with a three-dimensional $\cN=2$ effective action. 
A natural next step is to also investigate the F-theory uplift of our findings 
to a four-dimensional $\cN=1$ theory. This requires for the internal manifold $Y_4$
to be elliptically fibered. Shirking the fiber volume then yields the appearance 
of an extra circle. This limit is clearly complicated and requires the inclusion of further 
states that are not present in supergravity. However, applied to our reduction results the complications
are even more immediate. In fact, it is an interesting open question
how non-topological terms, for example including the warp-factor, are lifted to 
four space-time dimensions.

\vspace*{.5cm}
\noindent
\subsection*{Acknowledgments}

We would like to thank I\~naki Garc\'ia-Etxebarria, Daniel Junghans, Luca Martucci, Ruben Minasian, Diego Regalado, and 
Raffaele Savelli for useful discussions. This work was supported by a grant of the Max Planck Society. 



\end{document}